\documentclass[journal]{IEEEtran}

\usepackage{cite}
\usepackage{amsmath,amssymb,amsfonts}
\usepackage{graphicx,color}
\usepackage{textcomp}
\usepackage{orcidlink}
\def\BibTeX{{\rm B\kern-.05em{\sc i\kern-.025em b}\kern-.08em
    T\kern-.1667em\lower.7ex\hbox{E}\kern-.125emX}}
\AtBeginDocument{\definecolor{ojcolor}{cmyk}{0.93,0.59,0.15,0.02}}
\def\OJlogo{\vspace{-4pt}\hskip-4pt\includegraphics[height=18pt]{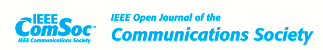}}

\usepackage{standalone}

\usepackage{algorithm}
\usepackage{algorithmicx}
\usepackage{algpseudocode}
\usepackage{setspace}
\usepackage{multicol}
\usepackage{enumerate}
\usepackage[font=small, belowskip=-15pt,aboveskip=0pt]{caption}
\usepackage{cite}
\usepackage{amsthm}
\usepackage[]{graphicx}
\usepackage{tikz}
\usepackage{tkz-tab}
\usepackage{pgfplots}
\usepackage{booktabs}
\usepackage{floatflt}

\usepackage{psfrag}	
\usepackage{array}
\usepackage{multirow,hhline}
\usepackage{exscale}
\usepackage{color}
\usepackage{colortbl}
\usepackage{epsfig}

\usepackage{array}
\usepackage[tight]{subfigure}

\usepackage[latin1]{inputenc} 
\usepackage[T1]{fontenc} 
\usepackage{rotating}
\usepackage{pbox}

\usepackage{pifont}

\usepackage{comment}
\usepackage{bookmark}
\usepackage{cleveref}

\renewcommand{\vec}[1]{\ensuremath{\boldsymbol{#1}}}

\newtheorem{remark}{Remark}

\usetikzlibrary{shapes,snakes}
\usetikzlibrary{calc}
\usetikzlibrary{patterns}
\usetikzlibrary{decorations.pathmorphing} 
\usetikzlibrary{spy}
\usetikzlibrary{positioning}
\usetikzlibrary{arrows, decorations.markings}
\usetikzlibrary{shapes.multipart, shapes.geometric, fit, backgrounds}
\usetikzlibrary{spy}

\definecolor{mycolor1}{RGB}{64, 122, 5}
\definecolor{mycolor2}{RGB}{97, 170, 209}
\definecolor{mycolor3}{RGB}{189, 0, 66}
\definecolor{mycolor4}{rgb}{0.92900,0.69400,0.12500}%
\definecolor{mycolor5}{RGB}{230, 142, 0}
\definecolor{mycolor6}{RGB}{10, 10, 130}

\renewcommand{\appendixname}{APPENDIX}

\begin{document}

\title{Robust Communication Design in RIS-Assisted THz Channels}


\author{\IEEEauthorblockN{Yasemin Karacora\orcidlink{0000-0001-8337-5745} \IEEEmembership{(Graduate Student Member, IEEE)}, Adam Umra\orcidlink{0009-0004-2071-2897} \IEEEmembership{(Graduate Student Member, IEEE)}, and Aydin Sezgin\orcidlink{0000-0003-3511-2662} \IEEEmembership{(Senior Member, IEEE)}}\\
\IEEEauthorblockA{Institute of Digital Communication Systems, Ruhr University Bochum, Germany \\ Emails: \{yasemin.karacora, adam.umra, aydin.sezgin\}@rub.de}
\thanks{This work has received funding from the programme ``Netzwerke 2021'', an initiative of the Ministry of Culture and Science of the State of Northrhine-Westphalia, Germany, and was supported in part by the German Federal Ministry of Education
and Research (BMBF) in the course of the 6GEM Research Hub under grant 16KISK037. The sole responsibility for the content of this publication lies with the authors.\\
The preliminary version \cite{karacora2024intermittency} of this work was presented at ICC 2024 in Denver, CO, USA.}
}
\markboth{Robust Communication Design in RIS-Assisted THz Channels}{Karacora \textit{et al.}}

\maketitle

\begin{abstract}
Terahertz (THz) communication offers the necessary bandwidth to meet the high data rate demands of next-generation wireless systems. However, it faces significant challenges, including severe path loss, dynamic blockages, and beam misalignment, which jeopardize communication reliability. 
Given that many 6G use cases require both high data rates and strong reliability, robust transmission schemes that achieve high throughput under these challenging conditions are essential for the effective use of high-frequency bands.
 In this context, we propose a novel mixed-criticality superposition coding scheme for reconfigurable intelligent surface (RIS)-assisted THz systems. This scheme leverages both the strong but intermittent direct line-of-sight link and the more reliable, yet weaker, RIS path to ensure robust delivery of high-criticality data while maintaining high overall throughput. We model a mixed-criticality queuing system and optimize transmit power to meet reliability and queue stability constraints. Simulation results show that our approach significantly reduces queuing delays for critical data while sustaining high overall throughput, outperforming conventional time-sharing methods. Additionally, we examine the impact of blockage, beam misalignment, and beamwidth adaptation on system performance. These results demonstrate that our scheme effectively balances reliability and throughput under challenging conditions, while also underscoring the need for robust beamforming techniques to mitigate the impact of misalignment in RIS-assisted channels. 
\end{abstract}


\section{INTRODUCTION}
Driven by the ultra-high data rate demands of future 6G applications, migrating to higher frequency bands, particularly the Terahertz (THz) regime, has emerged as a promising solution. 
THz communication offers the extensive bandwidth needed to support data-intensive applications, such as extended reality (XR) and digital twins. However, it also faces significant challenges, including severe path loss, molecular absorption, and high penetration and reflection losses. 
To ensure sufficient signal strength in THz channels, narrow pencil beams are required to focus the transmit power directly at the receiver. However, these highly directional beams make THz channels vulnerable to blockages from moving obstacles, or even the user itself. Given the severe attenuation of reflected paths in THz bands \cite{kokkoniemi2019NLOS300}, blockages of the line-of-sight (LoS) path can cause significant disruptions. 
In addition, pencil-shaped beams are prone to misalignment, even due to micro-mobility of the user. These challenges are exacerbated in dynamic environments, where accurate channel estimation and blockage prediction become nearly impossible. Thus, the unique characteristics of THz communication create a highly unpredictable and dynamic channel, leading to intermittent connectivity \cite{saad2021sevenTHz, akyildiz2018combating}.

 Yet, many 6G applications, such as XR, require not only high data rates, but also strong reliability and low latency to ensure a seamless user experience, particularly for mission- and safety-critical use cases like healthcare and industrial automation. Consequently, the tradeoff between rate and reliability becomes a fundamental challenge for the effective utilization of high-frequency bands. 

 In order to mitigate blockages despite the lack of strong multipath components, reconfigurable intelligent surfaces (RIS) are regarded as a promising approach \cite{weinberger2023RIS, chaccour2020risk, zarini2023RISaided, aman2023downlink}. RIS-assisted channels provide an alternative transmission path that can be less affected by blockages through strategic RIS placement. By dynamically adjusting the phase shifts of the reflective elements, RIS can steer beams toward the receiver, achieving higher signal gain. 
 However, integrating RIS introduces additional complexities, particularly regarding precise beam alignment. The use of narrow beams formed by a large number of reflective elements necessitates substantial overhead for channel estimation, frequent phase reconfiguration, and synchronization between the base station and the RIS controller \cite{Wang2022RIS}. Hardware limitations such as discrete phase shifts and imperfections further complicate the implementation and effectiveness of RIS beam steering \cite{Wang2022RIS}. Consequently, the practical deployment of highly directive beams in RIS-assisted THz communication poses difficulties, making less directional beams a more practical option. Furthermore, wider beams can address potential misalignments as they enhance robustness against user movements and require less frequent reconfiguration, thereby helping to reduce overhead \cite{delbari2024far, tian2023variableBW}. However, this advantage comes at the cost of lower gain, making it essential to consider the combined effects of misalignment, beamwidth, and gain reduction when evaluating the effectiveness of RIS deployment. 
 As a result, while the RIS-assisted reflective path is often more reliable than the direct LoS path, it remains significantly weaker.  
 Therefore, this work addresses the challenge of effectively leveraging RIS for blockage mitigation in THz communication by accounting for these trade-offs between reliability and signal strength. 
 
\begin{figure}
    \centering
    \includegraphics[width=\linewidth]{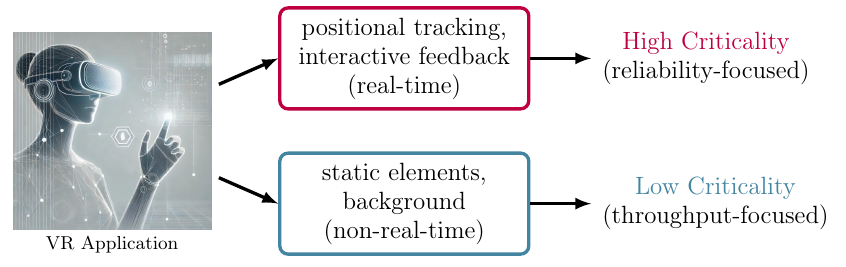}
    \caption{Example for data significance classification in a VR application, highlighting different QoS priorities for high- and low-criticality information.}
    \label{fig:VR}
\end{figure}
While such a tradeoff is typically addressed by the attempt of finding a compromise between the different objectives, e.g., data rate and outage probability, the paradigm shift in 6G towards goal-oriented communication schemes offers a new perspective. By taking account of the semantic meaning of the transmit data in the context of a specific task or goal, many applications typically comprise heterogeneous data that vary in criticality and significance. 
{For instance, virtual reality (VR) applications place extremely high demands on rate, reliability, and latency, yet the transmitted data can be categorized by varying levels of importance (see Fig. \ref{fig:VR}). Highly critical data includes real-time positional tracking and interactive feedback, which is essential for an immediate and responsive user experience. Meanwhile, less time-sensitive data, like non-real-time rendering of static components, can tolerate slight delays without significantly affecting the overall VR experience.}
This motivates a novel approach to handling quality of service (QoS) tradeoffs, namely leveraging service differentiation and prioritization techniques. Criticality-aware transmission schemes can guarantee strictly reliable delivery for particularly critical data segments, while still serving the overall throughput required by an application. This work explores the potential of such a differentiated approach for RIS-assisted THz communication, where data streams of high and low criticality arrive simultaneously and are processed in parallel using superposition coding (SC) to optimize both reliability and throughput.

\subsection{Prior Art}
THz communication systems face significant challenges due to severe path loss, susceptibility to blockage, and beam misalignment. The deployment of RIS has gained increasing attention as a potential solution to enhance coverage and mitigate blockages in LoS paths, as explored in \cite{liu2021RISopportunities, weinberger2023RIS, sivadevuni2023RIS}. 
RIS are especially promising in high-frequency bands, where channel intermittency and weak multipath components are prevalent, driving research into RIS design for mmWave and THz bands \cite{alexandropoulos2024reconfigurable, dash2022grapheneRIS}. The potential of RIS-assisted THz communication has been investigated in numerous studies \cite{chaccour2020risk, zarini2023RISaided, aman2023downlink, nor2024mobileBlockersRISTHz, papasotiriou2021THzRISMisalign, du2022THzRISMisalign}.
For instance, in \cite{chaccour2020risk}, RIS are deployed as THz base stations to enable a seamless user experience in virtual reality applications. 
In \cite{zarini2023RISaided}, RIS is applied in THz systems to serve users with diverse QoS needs, accounting for coexisting direct and reflective links. 
While blockage or misalignment effects are overlooked in \cite{zarini2023RISaided}, the impact of dynamic and self-blockage on achieving QoS demands in RIS-aided mmWave and THz networks is analyzed in \cite{nor2024mobileBlockersRISTHz}. The work in \cite{aman2023downlink} examines the coverage probability of RIS-aided THz networks, considering that users can be served via the direct link, the RIS link, or both simultaneously. 
Despite the strong potential of RIS to enhance coverage, their effective deployment in THz systems presents several key challenges, which remain unaddressed in \cite{zarini2023RISaided, nor2024mobileBlockersRISTHz, aman2023downlink}.
These include difficulties in channel state information (CSI) acquisition, beamforming optimization, and mitigating hardware impairments, which can significantly degrade the signal strength achievable through the RIS path \cite{Wang2022RIS}. 
In addition, highly directive beams in THz bands further compromise link reliability due to beam pointing errors, as shown in \cite{abdalla2023THzMisalign, kokkoniemi2020impact}. Although some works have examined the impact of misalignment on RIS-assisted THz channels, e.g., \cite{du2022THzRISMisalign, papasotiriou2021THzRISMisalign}, they do not sufficiently address how to exploit multiple transmission paths with varying gain and reliability characteristics.
Thus, studying the potential of leveraging RIS for THz communication requires the joint consideration of dynamic blockage, beam misalignment and path loss, all of which have a major impact on achievable rates and reliability.
Furthermore, many existing studies assume that RIS is used only when the LoS link is unavailable, e.g., \cite{papasotiriou2021THzRISMisalign, du2022THzRISMisalign, tian2023variableBW}. However, the highly fluctuating nature of THz channels makes blockages unpredictable. As such, these approaches do not offer robust solutions to manage highly intermittent links.\looseness-1 

In summary, while the direct LoS path suffers from significant intermittency, RIS can provide a more stable alternative link, though typically with lower signal gain, limiting its ability to meet high data rate demands. Consequently, jointly leveraging the strong but intermittent direct link and the more reliable but weaker RIS path is essential to fulfill the QoS requirements of future 6G applications \cite{karacora2024intermittency}.
Balancing rate and reliability despite the intermittency and path loss in THz channels, has been explored in several works, e.g., \cite{chaccour2020HRLLC, boulogeorgos2021directional, chaccour2020risk, karacoraTHzTCOM}.
However, the tradeoff between rate and reliability in THz communication remains largely unexplored from the perspective of data significance or criticality, which is essential for many 6G applications.

The need for developing criticality-aware communication schemes has been addressed in prior works, e.g., \cite{reifert2023comeback}, where the consideration of mixed-critical QoS levels has been shown to enhance resilience in resource management.
The use of SC to jointly transmit data with varying levels of criticality has been introduced in \cite{karacora2023rate}, where multi-connectivity is leveraged in uplink rate-splitting for the purpose of ultra-reliable low-latency communication.
Utilizing SC in THz communication has beed studied in \cite{sarieddeen2021terahertz}, considering the transmission of multiple streams in a point-to-point link, as well as in multi-user non-orthogonal multiple access (NOMA) scenarios, yet not for the purpose of mixed criticality transmissions. The application of SC for multi-resolution broadcasting has been proposed in \cite{choi2015multicast}, where two data streams of different priority are transmitted to two users in a multicast NOMA setting. Different from \cite{choi2015multicast}, our work leverages SC to deliver mixed-criticality data to a single user. While this concept is related to hierarchical modulation schemes used in image/video coding, e.g., as in \cite{arslan2011hierarchical}, our approach leverages SC for mixed-criticality data delivery in THz systems, aiming to fulfill stringent rate and reliability demands for critical 6G applications.

\subsection{Contribution}
\begin{figure*}[htb]
    \centering
    \includegraphics[width=\linewidth]{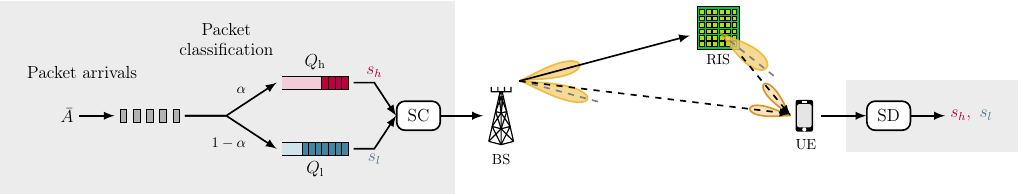}
    \caption{System model of a RIS-assisted BS-UE downlink channel. The direct BS-UE link as well as the RIS-UE channel are affected by dynamic blockage and beam misalignment. The arriving packets at the BS are classified according to their criticality and stored in a high- and a low-criticality (HC/LC) buffer. The BS applies superposition coding (SC) of the mixed-criticality data streams and the UE adopts a successive decoding (SD) approach.}
    \label{fig:sys_model}
\end{figure*}

This work aims to explore the potential of utilizing RIS in THz communication from a data significance perspective. We introduce a novel approach to address the trade-off between rate and reliability requirements, while accounting for the unique challenges of THz systems, including severe path loss, dynamic blockage, and beam misalignment. Specifically, we propose a framework that leverages weak multipath components, particularly a RIS-assisted reflective path, to mitigate the intermittency inherent in THz channels. By assuming that transmit data has different levels of criticality, we address the problem by providing enhanced reliability for high-criticality data using the RIS-path, while less critical data is delivered only when the direct LoS path is available, thereby increasing overall throughput. 
The key contributions can be summarized as follows:
\begin{itemize}
    \item We study RIS-assisted THz communication considering the impact of dynamic blockage and beam misalignment. Specifically, we examine downlink transmission through an intermittent LoS link and a more reliable but weaker RIS path (see Fig. \ref{fig:sys_model}).
    \item A novel approach to the rate-reliability tradeoff based on data significance is introduced. We assume that arriving data packets can be classified into high-criticality (HC) and low-criticality (LC) categories. The arrival of mixed-criticality data streams is modeled by a queuing system with two separate buffers. 
    \item We propose a mixed-criticality superposition coding (MC-SC) scheme, where both data streams are jointly transmitted via the RIS-assisted channel and successively decoded at the user. The proposed scheme ensures that HC data is reliably delivered as long as either the direct link or the RIS-path is available. In contrast, LC data is only decoded when the LoS path is unblocked. This approach meets reliability constraints for critical transmissions through leveraging path diversity while also maintaining high data rates by utilizing the stronger direct link. 
    \item We formulate a transmit power optimization problem in order to achieve mean queue stability for both data streams, while posing more stringent reliability constraints on the HC data. The non-convex optimization problem is solved iteratively, by applying successive convex approximation and fractional programming techniques.
    \item Simulations are conducted to evaluate throughput and outage probability for varying ratios of HC and LC data. We also analyze the impact of different blockage probabilities, beam pointing errors, and beamwidth adaptation on the rate-reliability tradeoff.
    \item Simulations of the queuing system demonstrate that the proposed scheme achieves significantly lower queuing delay and peak queue length for a substantial portion of HC data, while still meeting the required throughput. Furthermore, the SC scheme is shown to support higher portions of reliable HC transmission compared to a time-sharing approach.
\end{itemize}
The rest of this paper is organized as follows. Section \ref{sec:sysModel} presents the system model of the RIS-aided THz channel and introduces the proposed MC-SC scheme. In Section \ref{sec:problem}, the power allocation optimization problem is formulated. Section \ref{sec:results} discusses the simulation results, followed by a conclusion in Section \ref{sec:conclusion}.  

\emph{Notation:} Vectors are denoted by boldface letters. The operators $\mathbb{E}[\cdot]$, $|\cdot|$, and $\text{erf}(\cdot)$ represent the expectation, the absolute value and the error function, respectively. Additionally, $[x]^+$ denotes the positive part of $x$, i.e., $\max(0, x)$.

\section{SYSTEM MODEL}\label{sec:sysModel}
We consider a THz downlink transmission scenario, where a single base station (BS) transmits mixed-criticality data to a single user equipment (UE) via a direct LoS link and a RIS-enabled reflected path as illustrated in Figure \ref{fig:sys_model}. Both the BS and the UE are equipped with highly directive antennas, generating narrow beams. 
At the BS, the data packets intended for the user are classified and assigned to one of two distinct criticality levels, denoted as high-criticality (HC) and low-criticality (LC). These may correspond to different applications or services, which have heterogeneous reliability and latency requirements. Alternatively, within a goal-oriented communication framework, data packets from a single service can be categorized based on their significance (such as in a VR application as illustrated in Fig. \ref{fig:VR}). 
Here, we define $\alpha \in [0,1]$ as the fraction of data classified as HC.

\subsection{Channel Model}
Due to the severe attenuation induced by scattering in the THz band, we neglect the non-line-of-sight (NLoS) components and only consider the dominant LoS path (indicated as `direct path') and one RIS-aided reflection path (denoted as `RIS-path') in our channel model.
The channel coefficient of the direct link between BS and UE is given by
\begin{equation} \label{channel_coeff}
    {h} = \beta_d \eta_d \sqrt{\rho_d}, 
\end{equation}
where $\beta_d \in \{0,1\}$ is the random blockage variable, $\eta_d$ denotes the path gain coefficient, and $\rho_d$ captures the misalignment fading. 
Similarly, the channel coefficient of the RIS-path is defined as 
\begin{equation}
    g = \beta_r \eta_r \sqrt{\rho_r}.
\end{equation}
The specific models for blockage, path loss, and misalignment are described next.

\subsubsection{Blockage}
Due to the high penetration loss THz channels are vulnerable to dynamic blockage caused by other objects or the user itself. 
Therefore, THz channels suffer from a significant random intermittency that can lead to frequent outages. We assume that with an appropriate RIS placement, the BS-RIS link remains unaffected by such intermittency. Furthermore, since the RIS is positioned closer to the user than the BS, the RIS-assisted path is expected to be more reliable.
We model the intermittency of the direct LoS path and the RIS-path using Bernoulli random variables, $\beta_d$ and $\beta_r$, respectively. Each variable equals one when its respective link is available, and zero when the link is blocked.
We define blockage probabilities $q_d = \text{Prob}(\beta_d=0)$ and $q_r= \text{Prob}(\beta_r=0)$, with $q_r < q_d$.

 \subsubsection{Path Gain Coefficient}
 The channel gain of the direct link is comprising free space path loss and molecular absorption. It is given as \cite{aman2023downlink}
\begin{equation}
    \mathrm{\eta}_d = \frac{\sqrt{G_\mathrm{B} G_\mathrm{U}} c}{4\pi f d_\mathrm{BU}} e^{-\frac{1}{2} k_a(f) d_\mathrm{BU}}, 
\end{equation}
where $G_\mathrm{B}$ and $G_\mathrm{U}$ are the antenna gains of the BS and UE, $f$ is the operating frequency, and $c$ is the speed of light. The distance between the BS and the UE is denoted by $d_\mathrm{BU}$. The frequency-dependent molecular absorption coefficient is given by $k_a(f)$, which is obtained based on the model presented in \cite{kokkoniemi2021line}, valid for the frequency range of $100 - 450$ GHz. 

Similarly, the channel gain of the RIS path is modeled as
\begin{equation}
\eta_r = \frac{\sqrt{G_B G_R G_U}c}{4\pi f d_\mathrm{BR} d_\mathrm{RU}} e^{-\frac{1}{2}k_a(f) (d_\mathrm{BR}+d_\mathrm{RU})}.
\end{equation}
Here, $G_R$ denotes the gain of the RIS-reflected beam, while $d_\mathrm{BR}$ and $d_\mathrm{RU}$ are the distances between BS and RIS, and between RIS and UE, respectively.

\subsubsection{Misalignment Fading}
\begin{figure}
    \centering
    \includegraphics[width=0.7\linewidth]{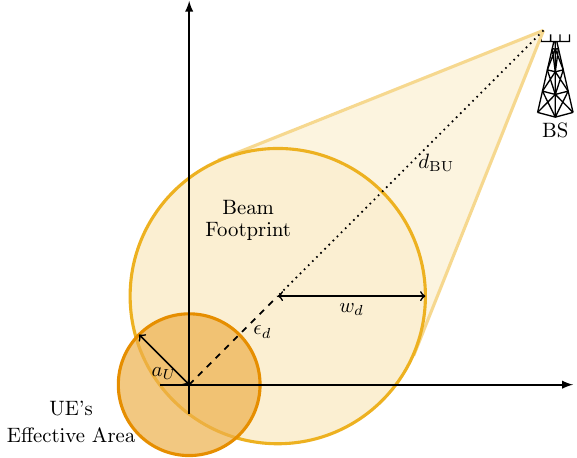}
    \caption{Illustration of beam misalignment on the direct BS-UE path, showing the transmission beam footprint and UE's effective area with a pointing error $\epsilon_d$.}
    \label{fig:misalignment}
\end{figure}
In addition, pencil-shaped beams entail a high risk for beam misalignment even due to micro-mobility of the user. We use a stochastic model to capture pointing errors as proposed in \cite{farid2007opticalPE} and widely used for highly directive THz beams \cite{boulogeorgos2019analytical, papasotiriou2020THzMisalignPN, abdalla2023THzMisalign, du2022THzRISMisalign, kokkoniemi2020impact, boulogeorgos2022rain}. The BS is assumed to transmit a highly directive circular beam via the direct path, which has a Gaussian shape with radius $w_d$ at the distance of the user, as shown in Fig. \ref{fig:misalignment}. The relation between antenna gain $G$ and beamwidth $w$ at distance $d$ for a Gaussian-shaped beam is given by \cite{stratidakis2021RISbeam}
\begin{equation} \label{gain_BW_relation}
    G = \frac{8 d^2}{w^2}.
\end{equation}
From this, we derive the beamwidth at the user as $w_d = \sqrt{8} d_\mathrm{BU}/\sqrt{G_B}$.
The UE has a circular detection beam of radius $a_U$. The pointing error displacement at the receiver, i.e., the radial distance between the centers of transmission and reception beams, is denoted by $\epsilon_d$. The misalignment fading coefficient, which indicates the fraction of power collected by the UE via the direct LoS link is approximated by \cite{farid2007opticalPE}
\begin{equation} \label{rho_d}
    \rho_d \approx A_d \exp\left(-\frac{2 \epsilon_d^2}{w_\mathrm{eq,d}^2}\right). 
\end{equation}
Here, $A_d$ is the fraction of power collected by the user via the direct path in case of perfect beam alignment, i.e., $\epsilon_d=0$, and is given as
 \begin{equation}\label{A_d}
     A_d = \text{erf} (v_d)^2,
 \end{equation}
 where
 \begin{equation}\label{v_d}
     v_d = \frac{\sqrt{\pi} a_U}{\sqrt{2}w_d}.
 \end{equation}
 In \eqref{v_d}, $a_U$ is the radius of the effective area of the reception antenna, given by \cite{boulogeorgos2022rain}
 \begin{equation}
     a_U = \frac{c \sqrt{G_U}}{2 \pi f}.
 \end{equation}
 The equivalent beamwidth $w_\mathrm{eq,d}$ (as defined in \cite{farid2007opticalPE}) is obtained from $w_d$ via the following relation:
 \begin{equation}\label{w_eq}
     w_\mathrm{eq,d}^2 = w_d^2 \frac{\sqrt{\pi} \text{erf}(v_d)}{2v_d \exp(-v_d^2)}.
 \end{equation}
We assume that $\epsilon_d$ follows a Rayleigh distribution with variance $\sigma_{m,d}^2$. Then, the probability density function (PDF) of $\rho_d$ can be written as \cite{farid2007opticalPE, boulogeorgos2022rain}
\begin{equation} \label{pdf_rho_d}
    f_{\rho_d} (x) = \frac{\gamma_d^2}{A_d^{\gamma_d^2}} x^{\gamma_d^2-1}, \quad 0 \leq x \leq A_d,
\end{equation}
in which 
\begin{equation} \label{gamma_d}
    \gamma_d = \frac{w_{eq,d}}{2 \sigma_{m,d}}.
\end{equation}
Hence, the cumulative distribution function (CDF) follows from \eqref{pdf_rho_d} as
\begin{equation} \label{cdf_rho_d}
    F_{\rho_d} (x) = \left(\frac{x}{A_d}\right)^{\gamma_d^2}, \quad 0 \leq x \leq A_d.
\end{equation}

For the RIS-path, we assume that the transmit beam of the BS to the RIS is perfectly aligned, given the static placement of BS antenna and RIS. However, the reflective beam of the RIS is subject to random pointing errors, modeled by the displacement $\epsilon_r$. The reflected beam is assumed to have a circular shape with radius $w_r$ at the UE. Thus, based on the previously described misalignment fading model \cite{farid2007opticalPE}, the fraction of collected power via the RIS-path is expressed as 
\begin{equation}\label{rho_r}
\rho_r \approx A_\mathrm{RIS} A_r \exp\left(-\frac{2 \epsilon_r^2}{w_\mathrm{eq,r}^2}\right).
\end{equation}
In \eqref{rho_r}, $A_r$ is the maximum power fraction collected via the RIS-UE link with perfect alignment, and $w_\mathrm{eq,r}$ denotes the equivalent width of the reflected beam. These parameters for the RIS-UE path are derived analogously to the direct path, following equations \eqref{A_d} -- \eqref{w_eq}. The fraction of power collected by the RIS from the BS beam is represented by $A_\mathrm{RIS}$ and obtained similarly to \eqref{A_d} -- \eqref{v_d}. Here, the width of the BS beam on the RIS surface is obtained from the relation \eqref{gain_BW_relation} and the radius of the RIS effective area is approximated as $a_\mathrm{RIS} = \frac{\lambda}{4} \sqrt{N_R}$, assuming a RIS with $\sqrt{N_R} \times \sqrt{N_R}$ elements of size $\lambda/2$.
For tractability, we neglect any correlation between pointing errors of the two propagation paths, assuming that the angles of arrival from the RIS and the BS are sufficiently distinct. Thus, $\epsilon_r$ is independent of $\epsilon_d$ and follows a Rayleigh distribution with variance $\sigma_{m,r}^2$. Hence, the distribution of $\rho_r$ is analogous to \eqref{pdf_rho_d} and \eqref{cdf_rho_d}, with parameter $\gamma_r = \frac{w_\mathrm{eq,r}}{2\sigma_{m,r}}$.
As previously discussed, beam training and phase configuration at the RIS are particularly challenging \cite{Wang2022RIS}. Due to less frequent beam adjustments, the RIS beam may experience greater misalignment variance than the direct beam. Therefore, in our model we assume $\sigma_{m,r} > \sigma_{m,d}$. Furthermore, we assume that the reflected beam is wider than the direct beam due to a coarser codebook design and potential hardware limitations. 

Channel uncertainty at the BS arises from the blockage state, represented by $\vec{\beta}= [\beta_d, \beta_r]$, and the pointing error displacement, given by $\vec{\epsilon} = [\epsilon_d, \epsilon_r]$, which affect both the direct path and the RIS link. We assume that the BS has only statistical knowledge of $\vec{\beta}$ and $\vec{\epsilon}$, characterized by the blockage probabilities $q_d$ and $q_r$, as well as the misalignment variances $\sigma_{m,d}^2$ and $\sigma_{m,r}^2$. Our goal is to ensure robust transmission despite these uncertainties, which are particularly detrimental in (sub-)THz bands.

\subsection{Mixed Criticality Superposition Coding}

\begin{figure}[]
    \centering
    \subfigure[no blockage of direct path]{\label{fig:no_blockage}
    \includegraphics[width=0.9\linewidth]{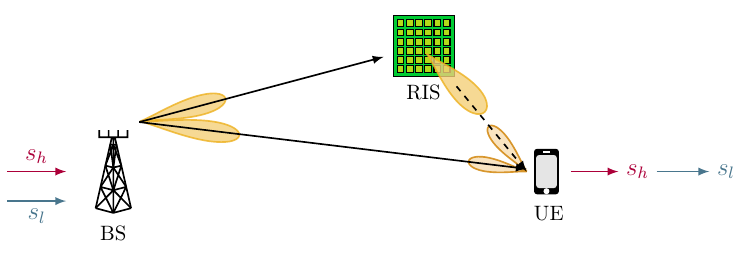}}
    \subfigure[direct path blocked]{\label{fig:with_blockage}
    \includegraphics[width=0.9\linewidth]{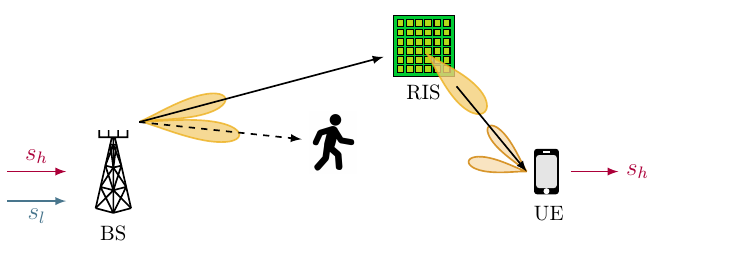}}
    \caption{Illustration of mixed-criticality transmission scheme: (a) As long as the direct path is available, both messages are successively decoded at the UE. (b) If the direct path is blocked, only the HC message is decoded via the RIS-path.}
    \label{fig:blockage}
\end{figure}

In our proposed scheme, SC is applied to the mixed-criticality data, allowing simultaneous transmission over the same frequency resources, while the user applies a successive decoding strategy. Here, higher transmit power is allocated to the HC data, enabling it to be decoded even under weaker channel conditions. In contrast, the LC data is successively decoded only when the channel is sufficiently strong. When applied to a RIS-aided THz channel with LoS intermittency, our scheme allows the decoding of the critical stream even if either one of the links is blocked, whereas the non-critical data remains decodable only when the LoS is available. This strategy is illustrated in Fig. \ref{fig:blockage}. Furthermore, with more power allocated to the HC stream and by leveraging path diversity, critical data experiences fewer outages caused by beam misalignment. 

\begin{remark}
Our MC-SC scheme is conceptually related to power-domain NOMA (e.g., \cite{islam2017NOMAsurvey}), which serves multiple users by superimposing signals at different power levels, enabling both a near (strong) and a far (weak) user to share the same channel. In our single-user design, the user similarly adapts to channel conditions: when the direct LoS path is available, both messages are decoded, but when the LoS is blocked, the UE acts as a 'weak user', decoding only the critical message. This approach enhances the robustness of critical message transmission against unexpected and unpredictable deep fades. 
\end{remark}

In the proposed scheme, the HC and LC messages, denoted by $s_h$ and $s_l$, respectively, are encoded and transmitted using two separate beams directed at the UE and the RIS.
We assume that the Angles of Departure (AoD) for the UE and RIS directions differ significantly compared to the narrow beamwidths and side lobes are negligible. This spatial separation allows for perfect isolation of the direct path and the RIS path. Accordingly, we define separate transmit signals for the beams directed at the UE and RIS as $x_d$ and $x_r$, respectively. By applying SC to the HC and LC messages, the transmit signals are given by: 
\begin{align}
    x_d &= \sqrt{p_h^{(d)}} s_h + \sqrt{p_l^{(d)}} s_l, \label{SC_x1}\\
    x_r &= \sqrt{p_h^{(r)}} s_h + \sqrt{p_l^{(r)}} s_l, \label{SC_x2}
\end{align}
where $p_h^{(d)}$ ($p_l^{(d)}$) and $p_h^{(r)}$ ($p_l^{(r)}$) are the transmit power allocated to the HC (LC) message in the direction of the user and the RIS, respectively. The received signal at the UE is then expressed as 
\begin{equation}
    y = h x_d + g x_r + n,
\end{equation}
in which $n \in \mathcal{CN}(0,\sigma_n^2)$ is additive white Gaussian noise (AWGN).

The user successively decodes both messages, starting with the HC message.
Thus, the signal-to-interference-plus-noise ratio (SINR) of the HC message can be written as
\begin{equation}
\label{sinr_h}
    \Gamma_h(\vec{\beta}, \vec{\epsilon}) = \frac{|h|^2 p_h^{(d)} + |g|^2 p_h^{(r)}}{|h|^2 p_l^{(d)} + |g|^2 p_l^{(r)} + \sigma_n^2},
\end{equation}
where the channel gains $|h|^2$ and $|g|^2$ of the direct and reflective path are determined based on \eqref{channel_coeff} with the blockage and misalignment states given by $\vec{\beta}$ and $\vec{\epsilon}$.
After correctly decoding and subtracting the HC message from the received signal, the LC data can be decoded, whereby the signal-to-noise-ratio (SNR) is given by
\begin{equation}
\label{sinr_l}
    \Gamma_l(\vec{\beta},\vec{\epsilon}) =\frac{|h|^2 p_l^{(d)} + |g|^2 p_l^{(r)}}{\sigma_n^2}.
\end{equation}
As a result, the achievable data rates with bandwidth $B$ are obtained as
\begin{align}
    r_h &= B \log_2(1 + \Gamma_h(\vec{\beta},\vec{\epsilon})),\\
    r_l &= B \log_2(1 + \Gamma_l(\vec{\beta},\vec{\epsilon})).
\end{align}
Let $R_h$ and $R_l$ denote HC and LC target rates and $\xi_h$ and $\xi_l$ be binary variables that are equal to one to indicate successful decoding of the HC and LC message, respectively, and equal to zero in case of an outage. Then, we have
\begin{equation}
    \xi_h = \begin{cases}
        1, & \text{if } r_h \geq R_h,\\
        0, & \text{if } r_h < R_h.
    \end{cases}
\end{equation}
As the decoding of the LC message depends on successful interference cancellation of the HC message, we have
\begin{equation}
    \xi_l = \begin{cases}
        1, & \text{if } \xi_h = 1 \text{ and } r_l \geq R_l,\\
        0, & \text{otherwise.}
    \end{cases}
\end{equation}

\subsection{Queuing Model}
As shown in Fig. \ref{fig:sys_model}, we adopt a queuing model at the BS, where packet arrivals are assumed to follow a Poisson process with arrival rate $\Bar{A}$. The arriving packets are then classified according to their criticality level, whereby a fraction of $\alpha$ data packets is considered HC data. We model the evolution of the two data queues containing HC and LC packets as follows:
\begin{align}
    Q_\mathrm{h}(t) &= \left[Q_\mathrm{h}(t-1) - \xi_h \frac{T}{M}R_\mathrm{h}(t) \right]^+ + \alpha A(t),\\
    Q_\mathrm{l}(t) &= \left[Q_\mathrm{l}(t-1) - \xi_l \frac{T}{M}R_\mathrm{l}(t) \right]^+ + (1-\alpha) A(t).
\end{align}
Here, $Q_\mathrm{h}(t)$ and $Q_\mathrm{l}(t)$ are the number of buffered HC and LC packets, and $A(t)$ represents the total packet arrivals in time slot $t$, respectively. The packet size is denoted by $M$, and $T$ is the time slot duration.
To avoid buffer overflow, the transmission scheme must ensure that all queues remain stable. A queue is considered stable if
\begin{equation}
    \lim_{t\rightarrow \infty} \sup \frac{1}{t} \sum_{\tau=0}^{t-1} \mathbb{E} [Q(\tau)] < \infty.
\end{equation}
In the long term, mean rate stability is achieved when the average departure rate exceeds the average arrival rate \cite{neely2010introduction}.

According to Little's law \cite{littlesLaw}, the average queuing delay, i.e., the mean waiting time of HC and LC data in each buffer, is given as 
\begin{equation}\label{delay}
    \tau_h = \frac{\mathbb{E}\{Q_h\}}{\alpha \bar{A}} \quad \text{and} \quad \tau_l = \frac{\mathbb{E}\{Q_l\}}{(1-\alpha) \bar{A}}.
\end{equation} 

Next, we formulate a power allocation problem to ensure mean queue stability while applying more stringent reliability constraints to the HC data.

\section{PROBLEM FORMULATION} \label{sec:problem}

Our goal is to serve the throughput demand specified by the packet arrival rate $\bar{A}$, while ensuring high reliability for the data classified as HC despite the fluctuating THz channel. More precisely, with the proposed SC scheme defined in \eqref{SC_x1} and \eqref{SC_x2}, we require the HC data to be decodable via the RIS if the direct link is blocked and vice versa. Hence, while the LC data stream suffers from outages whenever the direct LoS path is blocked, the HC data stream is disrupted only when both the LoS and the RIS-link become unavailable (see Fig. \ref{fig:blockage}). 
Additionally, we aim to stabilize connectivity in the presence of beam misalignment errors. Creating path diversity by leveraging RIS partly mitigates outages caused by spatial jitter and inaccurate beam configurations. To ensure robustness against small pointing errors, we adopt a heuristic approach, where decoding is successful via the direct link (RIS-link) as long as $\rho_d \geq \frac{1}{2}A_d$ ($\rho_r \geq \frac{1}{2}A_\mathrm{RIS} A_r$). Thereby, a link is assumed to become unavailable for beam displacements that lead to a power reduction exceeding $50\%$ at the receiver (which aligns with other works assuming outages to occur for beam misalignment beyond the half-power beamwidth). 
Thus, we formulate rate constraints based on the half-power beam gain, which occurs at the pointing error threshold $\vec{\epsilon}_{th} = \sqrt{\ln(\sqrt{2})} [w_\mathrm{eq,d}, w_\mathrm{eq,r}]$, derived from \eqref{rho_d} with $\rho_d=\frac{1}{2} A_d$. 
Thereby, the misdetection probability via the direct path caused by misalignment is obtained from \eqref{cdf_rho_d} as 
\begin{equation}
    q_{m,d} = F_{\rho_d}\left(\frac{A_d}{2}\right) = \left(\frac{1}{2}\right)^{\gamma_d^2}.
\end{equation} 
Analogously, the misdetection probability due to misalignment on the RIS-path follows as $q_{m,r} = \left(\frac{1}{2}\right)^{\gamma_r^2}$.

The outage probabilities can be derived from the blockage and misalignment probabilities as follows. Since LC data transmission relies solely on the availability of the direct path, it can be decoded only when this path is unblocked and the pointing error $\epsilon_d$ remains within the acceptable threshold.
 Thus, the outage probability for LC data is given by 
\begin{equation}
    P_\mathrm{out,l} = \text{Prob}(\xi_l=0) = 1-(1-q_d)(1-q_{m,d}).
\end{equation}
 For HC data, an outage occurs only if both the direct and RIS path fail to support decoding due to either blockage or misalignment. Note that in case of misalignment, decoding could be successful using the combined signal from both paths, even if individual paths do not provide sufficient signal strength. However, for simplicity, we approximate the outage probability by treating the paths independently, leading to
 \begin{equation}
 \begin{split}
    &P_\mathrm{out,h} = \text{Prob}(\xi_h=0)\\ &\quad\approx \left(1-(1-q_d)(1-q_{m,d})\right)\cdot\left(1-(1-q_r)(1-q_{m,r})\right).
    \end{split}
\end{equation}

We formulate an optimization problem to find the optimal power allocation to stabilize both (HC and LC) queues for a given $\alpha$. 
Note that mean queue stability is achieved when the average service rate is greater than the arrival rate. We introduce the optimization variable $\vec{\delta} = [\delta_h, \delta_l]$ denoting the gap between the rate of successfully delivered packets and the packet arrival rate. As optimization objective, we maximize the minimum of the weighted gap variables to ensure fairness for both queues.

Hence, defining $\vec{p} = [p_h^{(d)}, p_h^{(r)}, p_l^{(d)}, p_l^{(r)}]$ and $\vec{R} = [R_h, R_l]$, our optimization problem is formulated as follows:
\begin{subequations} \label{opt_original}
\begin{align}
\max_{\vec{\delta}, \vec{p}, \vec{R}} ~& \min\{\delta_h, \delta_l\}\tag{\theparentequation}\\
\text{s.t.}~~ &(1-P_{\mathrm{out},h})\frac{T}{M} R_h - \alpha \bar{A}\geq \alpha \delta_h,  \label{HC_stability_constr}\\
 & (1-P_{\mathrm{out},l})\frac{T}{M} R_l - (1-\alpha)\bar{A} \geq (1-\alpha) \delta_l,  \label{LC_stability_constr} \\
 & \delta_h, \delta_l \geq 0, \label{delta_constr} \\
\begin{split} &R_h \leq B \log_2\left(1+\Gamma_{h}(\vec{\beta}, \vec{\epsilon}_{th})\right),\\ & \hspace{60pt} \quad \forall \vec{\beta}\in \{(0,1), (1,0), (1,1)\},\end{split}\label{HC_rate_constr1}\\
&R_l \leq  B \log_2\left(1+\Gamma_{l}(\vec{\beta}, \vec{\epsilon}_{th})\right), \qquad \beta_d=1,\label{LC_rate_constr}\\
 & p_h^{(d)} + p_h^{(r)} +p_l^{(d)} + p_l^{(r)} \leq P_\mathrm{max}. \label{power_constr}
 \end{align}
\end{subequations}
Here, the constraints \eqref{HC_stability_constr}, \eqref{LC_stability_constr}, and \eqref{delta_constr} provide mean stability of the HC and LC queue. Meanwhile, \eqref{HC_rate_constr1} ensures that the HC data stream can be decoded as long as at least one of the two paths are available. The LC data stream, however, is only required to be decodable if the direct link exists as given by \eqref{LC_rate_constr}. Finally, \eqref{power_constr} is the transmit power constraint of the BS. 
Note that as the transmission of LC data solely relies on the direct LoS path, the optimal power allocation leads to ${p_l^{(r)}}^*=0$, i.e., the beam transmitted towards the RIS only contains a HC signal component.   

The problem \eqref{opt_original} is non-convex due to the rate expressions. Thus, we adopt a successive convex approximation (SCA) approach to iteratively optimize the power allocation. Based on a fractional programming framework proposed in \cite{shen2018fractional}, the original problem is approximated by a convex problem via a quadratic transform that is applied to the fractional SINR expressions. The detailed problem reformulation and the iterative algorithm are given in the Appendix.

\section{NUMERICAL RESULTS} \label{sec:results}
\begin{table}[tb]
    \centering
    \small{
    \begin{tabular}{|c|c|}
    \hline
      Transmit power $P_\mathrm{max}$   & 10 dBm \\ \hline
      Noise power $N_0$  &  -174 dBm/Hz\\ \hline
      Antenna gain $G_\mathrm{B}, G_\mathrm{U}$ & 40 dB, 35 dB\\ \hline
     Bandwidth $B$ & 10 GHz  \\ \hline
     Distances $d_\mathrm{BU}, d_\mathrm{BR}, d_\mathrm{RU}$ & 15 m, 15.8 m, 5 m  \\ \hline
     Carrier frequency $f$ & 300 GHz  \\ \hline
     Mol. absorption coeff. $k_a(f)$ & 0.0012 $\text{m}^{-1}$  \\ \hline
     RIS elements $N_\mathrm{R}$ &  $200\times 200$  \\ \hline
     Blockage prob. $q_d$, $q_r$ & 0.3, 0.1 \\ \hline
     Misalignment standard deviation $\sigma_{m,d},~\sigma_{m,r}$ & 0.1, 0.2 \\ \hline 
     RIS reflected beamwidth at UE $w_r$ & 0.8 m\\ \hline 
    \end{tabular}}
    \caption{Simulation Parameters (if not stated otherwise).}
    \label{tab:parameters}
\end{table}
The performance of our proposed MC-SC scheme is evaluated through numerical simulations with the parameters given in Table \ref{tab:parameters}.

\subsection{Achievability Analysis}
\begin{figure}[t]
    \centering
    \includegraphics[width=0.9\linewidth]{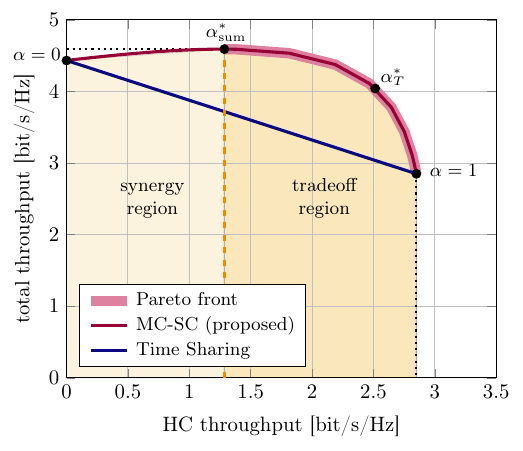}
    \caption{Feasibility region of the proposed MC-SC scheme in comparison to time sharing, showing the total achievable throughput versus the achievable HC throughput for $\alpha \in [0,1]$. The overall throughput is maximized at $\alpha_\mathrm{sum}^*$. For $\alpha < \alpha_\mathrm{sum}^*$, total throughput and the portion of reliable HC rate have a synergetic relation, whereas a tradeoff is exhibited for $\alpha > \alpha_\mathrm{sum}^*$. All points on the thick red line are Pareto optimal, and a tradeoff solution balancing throughput and reliability by solving \eqref{alpha_opt} is achieved with $\alpha_T^*$.}
    \label{fig:pareto}
\end{figure}

In the following, we explore the tradeoff between achievable throughput and reliability with our proposed scheme. First, we analyze the feasibility region of \eqref{opt_original} in Fig. \ref{fig:pareto}. To this end, we evaluate the maximum feasible $\bar{A}$, denoted by $\bar{A}_\mathrm{max}$, representing the achievable throughput, for different values of $\alpha \in [0,1]$. Note that at the border of the feasibility region, we obtain $\delta_h=\delta_l=0$.
Fig. \ref{fig:pareto} illustrates the feasible region of total throughput and the corresponding portion of HC throughput achieved with our proposed MC-SC scheme. 
For $\alpha=0$, where all data is treated as LC and transmitted solely via the direct LoS link, a maximum throughput of $4.4$ bit/s/Hz is achieved. When $\alpha=1$, meaning all data is treated as HC, the achievable throughput decreases to $2.8$ bit/s/Hz. This reduction occurs because, for the HC data, the transmit power is divided between the direct path and the RIS-link, and the stricter constraint in \eqref{HC_rate_constr1} ensures successful decoding even when one of the paths is blocked. Consequently, the higher reliability required for HC data comes at the cost of lower transmission rate, which is not fully compensated by the reduced outage probability. 
We observe that the achievable rate region with SC is significantly larger compared to a time sharing approach. 
Additionally, as $\alpha$ increases, the total throughput initially rises, reaching a maximum at $\alpha_\mathrm{sum}^* = 0.28$. This synergistic behavior occurs because, although the direct path has higher channel gains, frequent blockages lead to outages that reduce throughput. By using SC and exploiting path diversity for HC data, outages are reduced, resulting in improved throughput even with a lower transmission rate. However, as $\alpha$ continues to increase, the overall throughput decreases due to greater reliance on the weaker RIS channel. This introduces a tradeoff between total throughput and HC rate for $\alpha \in [\alpha_\mathrm{sum}^*, 1]$. 
The Pareto-optimal points are highlighted by the thick red line in Fig. \ref{fig:pareto}. The optimal value of $\alpha$ should be selected within the range $[\alpha_\mathrm{sum}^*, 1]$, whereby the desired tradeoff depends on the specific application requirements. For further analysis of our scheme, we suggest an optimal tradeoff solution that maximizes the sum of the normalized total throughput and HC throughput. Specifically, we define
\begin{equation}\label{alpha_opt}
    \alpha_T^* = \arg \max_{\alpha} \frac{\bar{A}_\mathrm{max}(\alpha)}{\bar{A}_\mathrm{max}(\alpha_{sum}^*)} +  \frac{\alpha \bar{A}_\mathrm{max}(\alpha)}{\bar{A}_\mathrm{max}(\alpha=1)}.
\end{equation}
The solution to \eqref{alpha_opt} can be determined using one-dimensional search algorithms such as golden section search. In the simulation depicted in Fig. \ref{fig:pareto}, the tradeoff solution is found at $\alpha_T^*=0.62$. With this choice of $\alpha$, the HC throughput nearly doubles compared to the maximum throughput solution, while the total throughput is reduced by only $12\%$.

\begin{figure}
    \centering
    \includegraphics[width=0.9\linewidth]{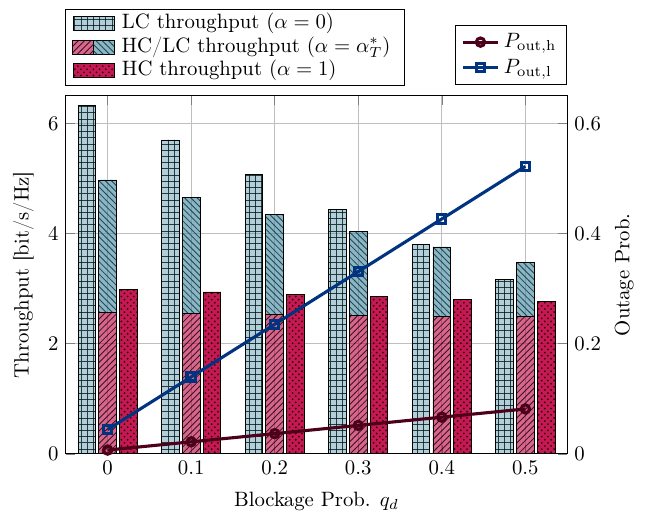}
    \caption{Throughput and outage probabilities of HC and LC transmission for different blockage probabilities of the direct path. The bars show the HC and LC throughput for $\alpha=0$, $\alpha_T^*$, and $\alpha=1$ (from left to right). The lines represent the corresponding outage probabilities.}
    \label{fig:rate_vs_q_d}
\end{figure}
Next, we analyze the impact of blockage probability and misalignment variance on the optimal tradeoff solution for $\alpha$.
Fig. \ref{fig:rate_vs_q_d} shows the achievable HC and LC throughput, as well as the corresponding outage probabilities, as functions of the direct path's blockage probability. We compare the throughput performance for three scenarios: (1) direct path only ($\alpha=0$), (2) our proposed MC-SC scheme with tradeoff parameter $\alpha_T^*$, and (3) treating all data as HC ($\alpha=1$). As the blockage probability increases, the outage probabilities for both HC and LC data rise. However, HC transmission is less impacted by direct path blockage due to the exploitation of the more reliable RIS path. As a consequence of less reliable transmissions, the throughput declines with increasing $q_d$. When transmitting solely via the direct path ($\alpha=0$), high throughput is achieved in the absence of blockage ($6.3$ bit/s/Hz), but it drops to $3.2$ bit/s/Hz when the direct link is available only $50\%$ of the time. In contrast, using both transmission paths ($\alpha=1$) provides greater robustness, with a lower rate that is less affected by the intermittency of the direct path. Therefore, while with $\alpha=1$, only $3$ bit/s/Hz are achievable even when $q_d=0$, the throughput decreases by just $7.4 \%$ when $q_d=0.5$. 
With the proposed MC-SC scheme and the tradeoff parameter set to $\alpha=\alpha_T^*$, a significant portion of HC data can be transmitted with higher reliability, while still maintaining high overall throughput. As the direct path becomes increasingly intermittent, total throughput decreases by $30\%$ from $5$ to $3.5$ bit/s/Hz, but HC throughput remains around $2.5$ bit/s/Hz.

It is important to note that when the direct path is reliable (i.e., $q_d=0$), the benefit of using the additional RIS-assisted path in terms of outage probability is rather low, while the loss in throughput is significant. However, for high $q_d$, the RIS-path greatly reduces outages with only a minor drop in throughput. Therefore, the difference between $P_\mathrm{out,h}$ and $P_\mathrm{out,l}$ should be considered when selecting $\alpha$. Thus, in case of a rather reliable direct path, a lower $\alpha$ should be chosen to achieve higher data rates, whereas with an unreliable direct link, prioritizing reliability over throughput by selecting a larger $\alpha$ is more beneficial. 

\begin{figure}
    \centering
    \includegraphics[width=0.9\linewidth]{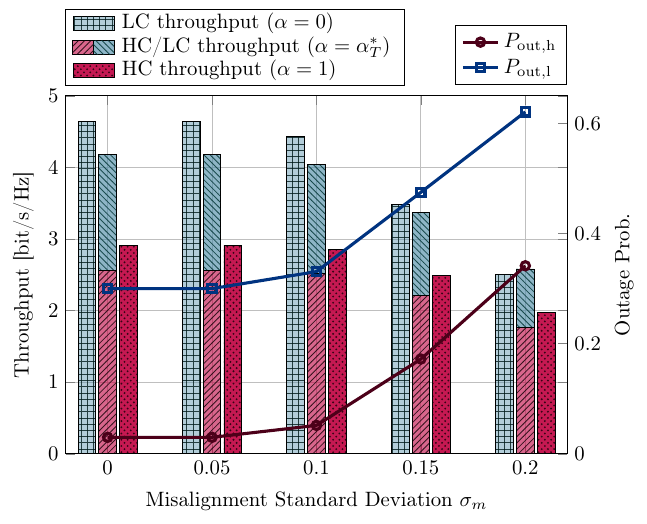}
    \caption{Throughput and outage probabilities of HC and LC transmission for different beam misalignment standard deviation. The bars show the HC and LC throughput for $\alpha=0$, $\alpha_T^*$, and $\alpha=1$ (from left to right). The lines represent the corresponding outage probabilities.}
    \label{fig:rate_sigma}
\end{figure}

Fig. \ref{fig:rate_sigma} shows the achievable throughput and corresponding outage probabilities for varying beam misalignment standard deviations. Here, we set $\sigma_{m,d} = \frac{1}{2} \sigma_{m,r} = \sigma_{m}$. Again, we compare the performance for $\alpha=0$, $\alpha_T^*$, and $\alpha=1$. As expected, outage probabilities increase significantly with greater misalignment (beyond $\sigma_m=0.1$), for both HC and LC transmissions. Although $P_\mathrm{out,h}$ remains much lower than $P_\mathrm{out,l}$, both exhibit similar growth as misalignment affects both links. This is reflected in the throughput, which decreases as $\sigma_m$ increases. Notably, LC throughput declines more rapidly due to the higher risk of outages. 
With the proposed MC-SC scheme and $\alpha=\alpha_T^*$, the achievable HC throughput is just slightly reduced compared to $\alpha=1$, yet the total throughput is substantially higher. For $\sigma_m=0.2$, the total throughput achieved with the tradeoff solution even exceeds the LC throughput with $\alpha=0$. As $\sigma_m$ increases, the value of $\alpha_T^*$ rises, since the detrimental effects of misalignment make the use of the additional RIS path more beneficial, despite the reduced channel gain. However, while a larger portion of data can be treated as HC with higher $\sigma_m$, the HC transmission also becomes less reliable, which may be unsuitable for the functionality of certain applications. This highlights the need for robustness measures to mitigate the impact of misalignment.

\begin{figure}
    \centering
    \includegraphics[width=0.9\linewidth]{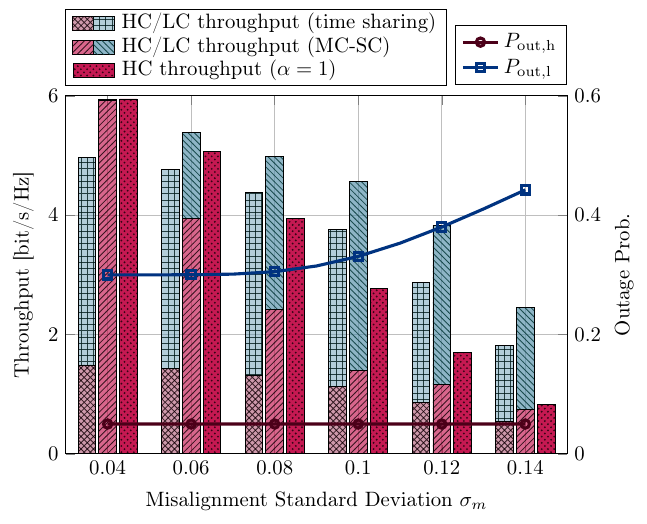}
    \caption{Throughput and outage probabilities of HC and LC transmission for different beam misalignment standard deviation in a scenario with strict HC requirements. The bars show the HC and LC throughput for a time sharing approach with $\alpha=0.3$, MC-SC with $\alpha \geq 0.3$, and the maximum HC throughput with $\alpha=1$ (from left to right). The lines represent the corresponding outage probabilities, whereby $P_\mathrm{out,h}=0.05$ is guaranteed for all values of $\sigma_m$.}
    \label{fig:rate_BW_sigma}
\end{figure}

Therefore, in Fig. \ref{fig:rate_BW_sigma}, we study a scenario with strict HC requirements. In this case, at least $30\%$ of the data must be transmitted as HC (i.e., $\alpha \geq 0.3$), with an outage probability of $P_\mathrm{out,h}=0.05$. As beam pointing errors increase, robustness measures are necessary to ensure the required reliability for HC data, e.g., through the employment of effective beamforming strategies. We apply a beamwidth adaptation approach to maintain a constant outage probability for HC transmission despite varying $\sigma_m$. Specifically, the beamwidth of the reflected beam $w_r$ is adjusted to meet the desired $P_\mathrm{out,h}$\footnote{Approaches for RIS beam design with tunable beamwidth have been proposed in \cite{delbari2024far, tian2023variableBW}.}. This comes at the cost of a reduced gain $G_R$, as given by \eqref{gain_BW_relation}. Note that widening the reflective beam improves the reliability of the RIS path, but also leads to higher path attenuation, thereby intensifying the disparity between the direct path and the RIS path. 
Fig. \ref{fig:rate_BW_sigma} shows the achievable HC and LC throughput, along with their corresponding outage probabilities, under these conditions. We compare three different strategies: (1) time sharing with $\alpha=0.3$, (2) the proposed MC-SC with $\alpha = \max(0.3, \alpha_\mathrm{sum}^*)$, and (3) treating all data as HC ($\alpha=1$). All three transmission strategies involve beamwidth adjustment and meet the given HC requirements. 

When beam pointing errors are small, the RIS can create a highly directive beam toward the UE without violating the required HC reliability constraints\footnote{Note that the beam focusing capability of the RIS is restricted by several factors, such as the number of reflective elements and the discrete phase shifter resolution, which are not considered in this simulation.}. This narrow beam results in higher gain, strengthening the RIS path. Consequently, at $\sigma_m=0.04$, the highest throughput is achieved when $\alpha=1$. However, as misalignment increases, the reflective beam must be widened to maintain the required $P_\mathrm{out,h}$, leading to a rapid decrease of HC throughput. Thus, imposing strict HC requirements on all data (i.e., $\alpha=1$) leads to a significant decline in overall throughput. Applying the data significance approach with time sharing, where only $30\%$ of the data must meet strict reliability standards, the total throughput decreases more slowly. However, the proposed MC-SC scheme outperforms both other strategies in terms of total throughput across the entire range of $\sigma_m$. Note that for $\sigma_m < 0.1$, we have $\alpha_\mathrm{sum}^* > 0.3$, meaning that more than the required $30\%$ of data is transmitted as HC. For $\sigma_m \geq 0.1$, $\alpha=0.3$ is chosen to meet the requirements, though this is not throughput-optimal. 
Nevertheless, with high impact of misalignment at $\sigma_m=0.14$, the MC-SC scheme nearly triples the total throughput compared to the scenario where all data is treated as HC ($\alpha=1$). Additionally, MC-SC provides a $35\%$ improvement in throughput over the time sharing approach. 

In summary, the results from Figs. \ref{fig:rate_sigma} and \ref{fig:rate_BW_sigma} demonstrate that while RIS is promising for mitigating dynamic blockages in THz communications, beam pointing errors must be accounted for. Applications with strict reliability requirements necessitate robust transmission strategies, such as beamwidth adaptation \cite{karacoraTHzTCOM}, to address spatial jitter and user micro-mobility. However, these techniques come at the cost of reduced beamforming gain, intensifying the tradeoff between rate and reliability. Therefore, transmission schemes that consider data significance, like the proposed MC-SC, are a promising solution to balance these conflicting application requirements.

\subsection{Queuing System Analysis}

\begin{figure}
    \centering
    \includegraphics[width=0.85\linewidth]{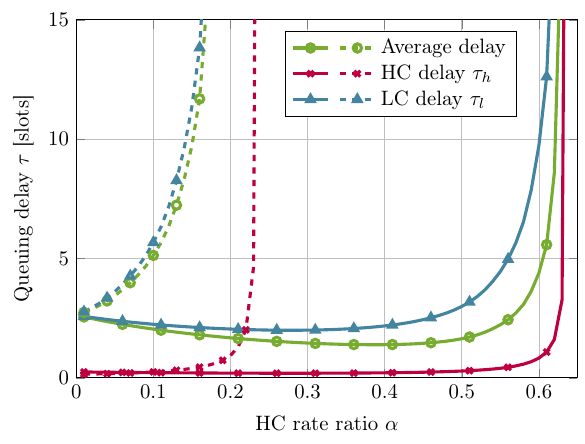}
    \caption{Average packet waiting time in the queue as a function of the HC rate ratio $\alpha$ for our proposed MC-SC scheme (solid lines) and a time sharing scheme as baseline (dashed lines).}
    \label{fig:delay}
\end{figure}

Next, we evaluate the average queuing delay, i.e., the mean waiting time of HC/LC data in each buffer, as obtained from \eqref{delay}. Fig. \ref{fig:delay} shows the average queuing delay for a packet size of $M=5$ Mbit, time slot duration $T=100$ ms, and a total packet arrival rate of $\bar{A}=800$ packets/slot. We compare the proposed MC-SC scheme with a time sharing approach, where a portion of each time slot is allocated to transmit HC data with enhanced reliability over both transmission paths, while the remaining time is used to transmit LC data through the direct link. 
When $\alpha=0$, the average (LC) queuing delay is approximately $2.5$ time slots. Due to the different reliability levels of HC and LC data, the HC delay achieved for sufficiently small $\alpha$ is much lower, approximately by a factor of 10. 
With time sharing, the LC delay increases as soon as some data is classified as HC ($\alpha >0$), whereas the HC delay grows more slowly. The system becomes unstable around $\alpha>0.18$. In contrast, our proposed scheme allows for a significantly larger portion of data to be transmitted as HC with short delays. In fact, the average queuing delay even exhibits a slight decrease, reaching a minimum at $\alpha=0.39$. This is a result of the gain in sum-throughput that can be achieved with superposition coding. The overall delays remain low when less than $50\%$ of the packets are treated as HC data, and the system becomes unstable at $\alpha> 0.63$.

\begin{figure}
    \centering
    \includegraphics[width=0.85\linewidth]{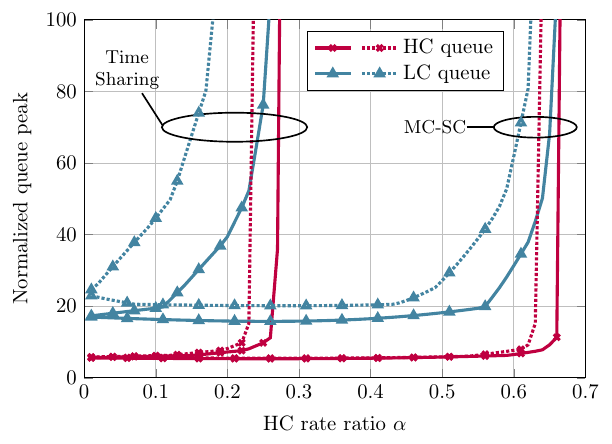}
    \caption{Normalized peak queue length of the HC and LC buffers as a function of the HC rate ratio $\alpha$ with/without misalignment (dotted/solid) for our proposed MC-SC scheme and a time sharing scheme as baseline.}
    \label{fig:peak}
\end{figure}

Fig. \ref{fig:peak} shows the normalized peak queue length, defined as $\max_{t} \frac{Q_\mathrm{h}(t)}{\alpha \bar{A}}$ and $\max_{t} \frac{Q_\mathrm{l}(t)}{(1-\alpha) \bar{A}}$ for the HC and LC buffer, respectively. Again, we compare the MC-SC scheme with the time-sharing approach across varying $\alpha$. Furthermore, the performance is evaluated with and without beam misalignment. 
Due to the higher outage probability associated with LC transmission, packets may accumulate in the LC buffer until successful delivery to the UE. As a result, the peak queue length of HC data remains significantly lower than that of the LC queue. 
The peak queue length follows a trend similar to the average delay, showing that with MC-SC, queues remain stable without considerable packet accumulation over a much larger range of $\alpha$. Under beam misalignment, both schemes experience increased buffer congestion, and the range of $\alpha$, for which queues are stabilized, is reduced. However, with MC-SC, the peak queue length stays stable for a broad range of $\alpha$ (up to 0.4) even in the presence of misalignment. Compared to time sharing, MC-SC increases the maximum $\alpha$ that ensures stabilized queues by approximately $2.6$ times in the absence of misalignment, and even by a factor of four under misalignment conditions. Furthermore, while LC queue length slightly increases under misalignment in that range, the HC queue peak is barely affected by pointing errors until $\alpha$ reaches approx. 0.55, indicating its robustness against misalignment. Across the stable $\alpha$ range, the HC queue peak is around three times lower than the LC peak under perfect beam alignment, and about four times lower when pointing errors occur.

As a result, the simulations demonstrate the effectiveness of our proposed criticality-aware scheme in achieving low delays for particularly critical data, while maintaining queue stability. Furthermore, with MC-SC, the portion of critical data can be substantially enhanced compared to a time sharing approach, without causing packet accumulation.

\section{CONCLUSION} \label{sec:conclusion}
This paper addresses key challenges in THz communication: the severe path loss and high sensitivity to blockage and beam misalignment, which lead to link intermittency and a fundamental rate-reliability tradeoff. We propose a mixed-criticality superposition coding scheme for downlink transmission that leverages both the highly intermittent LoS link and the more reliable but weaker RIS-assisted path. The scheme prioritizes highly critical data to ensure reliability under adverse conditions, while low-criticality data is transmitted opportunistically when the LoS path is available.
By modeling a data queuing system at the base station, we formulated a power allocation problem to ensure queue stability and solve it using an iterative SCA-based algorithm. Simulation results demonstrate that our approach allows reliable transmission for a significant portion of critical data by exploiting the RIS-enabled path diversity, while maintaining high overall throughput. As blockage and misalignment increase, the system relies more on the RIS for robust transmission, while better conditions allow for significant throughput gains by opportunistically transmitting additional data via the direct link.
Moreover, the results show that beam misalignment, which affects both transmission paths, can substantially reduce reliability for both HC and LC data. However, beamwidth adjustment in combination with SC allows the system to meet reliability requirements for critical data while outperforming time sharing and single stream approaches in terms of total throughput. Our proposed approach supports substantially higher critical data rates without increasing queuing delay or causing packet accumulation, demonstrating its clear advantage over time-sharing solutions.

\begin{appendix}
The non-convex power allocation problem can be solved using a SCA approach. We derive a convex approximation of \eqref{opt_original} as follows:

First, we introduce auxiliary variables $\vec{\gamma}=[\gamma_h, \gamma_l]$. Hence, we can rewrite problem \eqref{opt_original} as follows:
\begin{subequations} \label{opt2}
\begin{align}
\max_{\vec{\delta}, \vec{p}, \vec{R}, \vec{\gamma}} ~& \min\{\delta_h, \delta_l\}\tag{\theparentequation}\\
\text{s.t.}~~ & \eqref{HC_stability_constr}, ~\eqref{LC_stability_constr},~ \eqref{delta_constr}, ~\eqref{power_constr}, \\
&R_h \leq B \log_2\left(1+\gamma_h\right), \label{HC_rate_constr2}\\
&R_l \leq  B \log_2\left(1+\gamma_l\right),\label{LC_rate_constr2}\\
& \gamma_h \leq \Gamma_{h}(\vec{\beta}, \vec{\epsilon}_{th}),\quad \vec{\beta}\in\{(0,1), (1,0), (1,1)\},\\
& \gamma_l \leq \Gamma_{l}(\vec{\beta}, \vec{\epsilon}_{th}), \quad \beta_d=1.
\end{align}
\end{subequations}
Next, we adopt the quadratic transform proposed in \cite{shen2018fractional} to handle the fractional SINR expressions. By introducing the auxiliary variables $\vec{\mu} = [\mu_{h,(0,1)}, \mu_{h,(1,0)}, \mu_{h,(1,1)}, \mu_l]$ and applying the approach from \cite{shen2018fractional}, we obtain the functions
\begin{align}
\begin{split}
g_{h,\vec{\beta}}(\vec{p}, \vec{\gamma})& = \gamma_h \hspace{-2pt}- \hspace{-2pt}2\mu_{h,\vec{\beta}} 
\sqrt{\beta_d\eta_{d}^2 \rho_d(\epsilon_{th}) p_h^{(d)}\hspace{-3pt} +\hspace{-2pt} \beta_r\eta_{r}^2 \rho_r(\epsilon_{th}) p_h^{(r)}}\\ &\hspace{-10pt} + {\mu_{h,\vec{\beta}}}^2 \left(\beta_d \eta_{d}^2 \rho_d(\epsilon_{th}) p_l^{(d)}\hspace{-2pt} + \hspace{-2pt}\beta_r\eta_{r}^2 \rho_r(\epsilon_{th}) p_l^{(r)} + \sigma_n^2\right),
\end{split} \label{g_h}\\
\begin{split}
g_{l}(\vec{p}, \vec{\gamma})& = \gamma_l - 2\mu_l 
\sqrt{\eta_d^2 \rho_d(\epsilon_{th}) p_l^{(d)}}+ {\mu_l}^2 \sigma_n^2.
\end{split} \label{g_l}
\end{align}
The optimal $\vec{\mu}$ for fixed $\vec{p}$ and $\vec{\gamma}$ can be obtained by setting the derivatives of \eqref{g_h} and \eqref{g_l} to zero. Thus, we have
\begin{align}
{\mu_{h,\vec{\beta}}}^* &= \frac{\sqrt{\beta_d \eta_{r}^2 \rho_d(\epsilon_{th}) p_h^{(d)} + \beta_r \eta_{r}^2 \rho_r(\epsilon_{th}) p_h^{(r)}}}{\beta_d \eta_{r}^2 \rho_d(\epsilon_{th}) p_l^{(d)} + \beta_r\eta_r^2 \rho_r(\epsilon_{th}) p_l^{(r)} + \sigma_n^2}, \label{mu_h}\\
{\mu_l}^* &= \frac{\sqrt{ \eta_{d}^2 \rho_d(\epsilon_{th}) p_l^{(d)}}}{\sigma_n^2}. \label{mu_l}
\end{align}
As for constant $\vec{\mu}$, the functions \eqref{g_h} and \eqref{g_l} are convex in $\vec{p}$ and $\vec{\gamma}$, the optimization problem \eqref{opt2} can be approximated by a convex problem with fixed $\vec{\mu}$:
\begin{subequations} \label{opt_convex}
\begin{align}
\max_{\vec{\delta}, \vec{p}, \vec{R}, \vec{\gamma}} ~& \min\{\delta_h, \delta_l\}\tag{\theparentequation}\\
\text{s.t.}~~ & \eqref{HC_stability_constr}, ~\eqref{LC_stability_constr},~ \eqref{delta_constr}, ~\eqref{power_constr},~ \eqref{HC_rate_constr2},~ \eqref{LC_rate_constr2}, \\
& g_{h,\vec{\beta}}(\vec{p}, \vec{\gamma}) \leq 0, \quad \vec{\beta} \in \{(0,1), (1,0), (1,1)\},\\
& g_l(\vec{p}, \vec{\gamma}) \leq 0.
\end{align}
\end{subequations}
Thus, the optimal power allocation is obtained via an iterative SCA-based algorithm. That is, the approximated problem \eqref{opt_convex} is solved using a convex optimization solver such as CVX \cite{cvx}, and the auxiliary variables $\vec{\mu}$ are alternately updated following \eqref{mu_h} and \eqref{mu_l}. The algorithm is summarized in Alg. \ref{alg:MCSC_alg}.
\begin{algorithm}[]
\caption{Power Allocation for the MC-SC scheme}
  \begin{algorithmic}[1]
    \State Initialize $\vec{p}$
    \Repeat 
        \State Compute $\vec{\mu}$ based on \eqref{mu_h},  \eqref{mu_l}
        \State Solve \eqref{opt_convex} for fixed $\vec{\mu}$
    \Until{Convergence}
    \end{algorithmic}
    \label{alg:MCSC_alg}
\end{algorithm}
\end{appendix}

\bibliography{references}
\bibliographystyle{IEEEtran}

\begin{IEEEbiography}[{\includegraphics[width=1in,height=1.25in,clip,keepaspectratio]{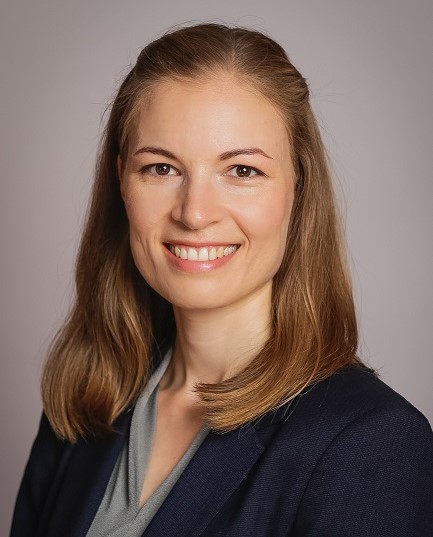}}]{YASEMIN KARACORA } (Graduate Student Member, IEEE) received the B.Sc. and M.Sc. degrees in electrical engineering and information technology from Ruhr University Bochum, Germany, in 2016 and 2019, respectively, where she is currently pursuing the Ph.D. degree with the Institute of Digital Communication Systems. From 2016 to 2017, she was an Exchange Student with the ECE Department, Purdue University, IN, USA. Her research interests include wireless communications, beamforming at (sub-)terahertz frequency bands, and reliability and resilience in 5G and 6G networks.
\end{IEEEbiography}

\begin{IEEEbiography}[{\includegraphics[width=1in,height=1.25in,clip,keepaspectratio]{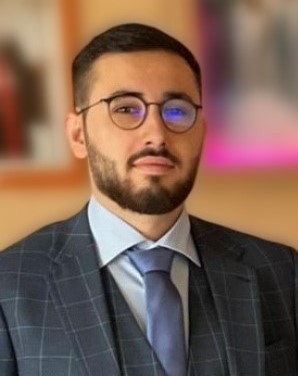}}]{ADAM UMRA } (Graduate Student Member, IEEE) received the B.Sc. and M.Sc. degrees in Electrical Engineering and Information Technology from Ruhr University Bochum, Germany, in 2021 and 2023, respectively. He is currently pursuing a Ph.D. at the Institute of Digital Communication Systems at the same university. His research focuses on wireless communications, MIMO radar signal processing, cognitive radar systems, and machine learning applications for integrated sensing and communication.
\end{IEEEbiography}

\begin{IEEEbiography}[{\includegraphics[width=1in,height=1.25in,clip,keepaspectratio]{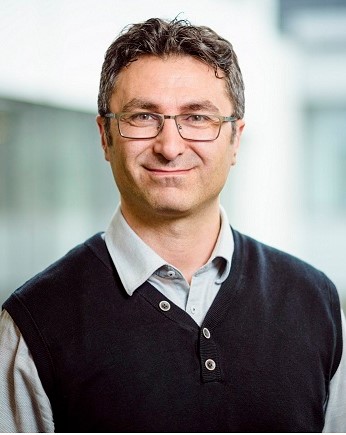}}]{AYDIN SEZGIN } (Senior Member, IEEE) received the Dr. Ing. (Ph.D.) degree in electrical engineering from TU Berlin, in 2005. From 2001 to 2006, he was with the Heinrich-Hertz-Institut, Berlin. From 2006 to 2008, he held a postdoctoral position, and was also a lecturer with the Information Systems Laboratory, Department of Electrical Engineering, Stanford University, Stanford, CA, USA. From 2008 to 2009, he held a postdoctoral position with the Department of Electrical Engineering and Computer Science, University of California, Irvine, CA, USA. From 2009 to 2011, he was the Head of the Emmy-Noether-Research Group on Wireless Networks, Ulm University. In 2011, he joined TU Darmstadt, Germany, as a professor. He is currently a professor with the Ruhr-Universität Bochum, Germany. He has published several book chapters, more than 70 journals and 200 conference papers in these topics. Aydin is a winner of the ITG-Sponsorship Award, in 2006. He was a first recipient of the prestigious Emmy-Noether Grant by the German Research Foundation in communication engineering, in 2009. He has coauthored papers that received the Best Poster Award at the IEEE Communication Theory Workshop, in 2011, the Best Paper Award at ICCSPA, in 2015, at ICC, in 2019, and at ISAP, in 2023. He was the General Co-Chair of the 2018 International ITG Workshop on Smart Antennas, the Program Co-Chair of 2019 Crowncom Conference, and the Workshop Co-Chair of the 2022 WCNC Workshop on Rate-Splitting and Next Generation Multiple Access. He has served as an Associate Editor for IEEE Trans. on Wireless Communications, from 2009 to 2014, an Area Editor for the Journal of Electronics and Communications, from 2010 to 2011, and is serving as Editor-in-Chief for Springer Wireless Personal Communications, since 2023.
\end{IEEEbiography}

\end{document}